%% file: ms.tex
\documentclass[6pt,onecolumn]{article}
\usepackage[english]{babel}

\usepackage{amsmath}
\usepackage{amsfonts}
\usepackage{amssymb}
\usepackage{graphicx}
\usepackage[affil-it]{authblk}
\usepackage{hyperref}
\usepackage{color}
\usepackage{caption}
\usepackage{verbatim}
\usepackage[normalem]{ulem}
\usepackage{subfigure}
\usepackage{slashed}
\usepackage{cancel}
\usepackage{braket}
\usepackage{multicol}
\usepackage{tcolorbox}
\tcbuselibrary{theorems}
\usepackage{bm}
\usepackage{enumitem}
\usepackage[numbers,sort&compress]{natbib}

\hypersetup{linktocpage=true}

\usepackage{tikz} 
\usepackage{tikz-feynman}

\tikzfeynmanset{compat=1.0.0}

\usetikzlibrary{arrows,shapes}
\usetikzlibrary{trees}
\usetikzlibrary{matrix,arrows} 				
\usetikzlibrary{positioning}				
\usetikzlibrary{calc,through}				
\usetikzlibrary{decorations.pathreplacing}  
\usepackage{pgffor}							

\usetikzlibrary{decorations.pathmorphing}	
\usetikzlibrary{decorations.markings}
\tikzset{
    vector/.style={decorate, decoration={snake}, draw},
	provector/.style={decorate, decoration={snake,amplitude=2.5pt}, draw},
	antivector/.style={decorate, decoration={snake,amplitude=-2.5pt}, draw},
    fermion/.style={draw=black, postaction={decorate},
        decoration={markings,mark=at position .55 with {\arrow[draw=black]{>}}}},
    fermionbar/.style={draw=black, postaction={decorate},
        decoration={markings,mark=at position .55 with {\arrow[draw=black]{<}}}},
    fermionnoarrow/.style={draw=black},
    gluon/.style={decorate, draw=black,
        decoration={coil,amplitude=4pt, segment length=5pt}},
    scalar/.style={dashed,draw=black, postaction={decorate},
        decoration={markings,mark=at position .55 with {\arrow[draw=black]{>}}}},
    scalarbar/.style={dashed,draw=black, postaction={decorate},
        decoration={markings,mark=at position .55 with {\arrow[draw=black]{<}}}},
    scalarnoarrow/.style={dashed,draw=black},
    electron/.style={draw=black, postaction={decorate},
        decoration={markings,mark=at position .55 with {\arrow[draw=black]{>}}}},
	bigvector/.style={decorate, decoration={snake,amplitude=4pt}, draw},
    line/.style={draw=black},
}\usetikzlibrary{decorations.markings}

\usepackage{fancyhdr}
\fancypagestyle{plain}{%
	\fancyhead[R]{\Large{TUM-HEP 1353/21}}
	}

\title{Singlet Scalars as Dark Matter and the Muon $(g-2)$ Anomaly}
\author[1]{Basti\'an D\'iaz S\'aez \thanks{bastian.diaz@tum.de}}
\author[2]{Karim Ghorbani\thanks{karim1.ghorbani@gmail.com}}
\affil[1]{Physik-Department, Technische Universität München, James-Franck-Straße, 85748 Garching, Germany}
\affil[2]{Physics Department, Faculty of Sciences, Arak University, Arak 38156-8-8349, Iran}

\begin{document}
\twocolumn
\maketitle
\begin{abstract}
We explore a simple model containing two singlet scalars
as dark matter candidates and a vector-like lepton.
Evading collider and dark matter constraints, the model is able to accommodate the correct dark matter relic abundance and the $(g-2)_\mu$ anomaly when the masses of the three new fields are around 100 GeV and the values of the minimum amount of required couplings are order one.
\end{abstract}


\input{1-Introduction}

\input{2-Model}
\input{3-Constraints}
\input{4-Analysis}

\input{5-Conclusions}





\bibliography{bibliography}
\bibliographystyle{utphys}

\end{document}

%% file: 1-Introduction.tex
\section{Introduction}\label{minimal_model}
The reported combined measurement of the anomalous magnetic moment of the muon $a_\mu \equiv \frac{g-2}{2}$ by BNL E821 \cite{Muong-2:2006rrc} and FNAL E989 \cite{Muong-2:2021ojo} experiments yields a 4.2$\sigma$ discrepancy from the Standard Model (SM) prediction \cite{Aoyama:2012wk,Aoyama:2019ryr,Czarnecki:2002nt,Gnendiger:2013pva,Davier:2017zfy,Keshavarzi:2018mgv,Colangelo:2018mtw,Hoferichter:2019gzf,Davier:2019can,Keshavarzi:2019abf,Kurz:2014wya,Melnikov:2003xd,Masjuan:2017tvw,Colangelo:2017fiz,Hoferichter:2018kwz,Gerardin:2019vio,Bijnens:2019ghy,Colangelo:2019uex,Blum:2019ugy,Colangelo:2014qya, Davier:2010nc, Aoyama:2020ynm}
\begin{eqnarray}\label{deltaa}
 \Delta a_\mu = a_\mu^\text{exp} - a_\mu^\text{SM} = (251 \pm 59)\times 10^{-11}.
\end{eqnarray}
This hint of new physics beyond the Standard Model (BSM) has motivated different extensions to the SM to account for it. In particular, new interactions involving the leptonic sector along with dark matter (DM) candidates seem an appealing and exiting avenue \cite{Belanger:2015nma,Kowalska:2017iqv, Barman:2018jhz, Calibbi:2018rzv,Kowalska:2020zve, Jana:2020joi, Chun:2020uzw, DeJesus:2020yqx, Athron:2021iuf, Arcadi:2021glq, Qi:2021rhh, Arcadi:2021cwg, Guedes:2021oqx, Lu:2021vcp, CarcamoHernandez:2021qhf, Hernandez:2021tii, Crivellin:2021rbq}. One of the simplest extensions to the SM to account for DM and the $(g-2)_\mu$ anomaly involves the introduction of two new fields, i.e. one real scalar plus one vector-like lepton (VLL), both odd under the same discrete symmetry $Z_2$ ($S_1\rightarrow -S_1$ and $\psi\rightarrow -\psi$). However, this set-up has been shown to give an under-abundant relic density \cite{Athron:2021iuf}. Extensions with three fields have also been explored \cite{Calibbi:2018rzv, Athron:2021iuf}, and in particular,
it has been shown that extending the minimal scenario (singlet scalar plus a VLL) by including a $SU(2)_L$ scalar doublet may accommodate $(g-2)_\mu$ and DM constraints, provided that the scalar DM masses be above 500-600 GeV, and in the TeV scale when the mixed scalars are degenerated \cite{Athron:2021iuf}. 

In this work we explore a simpler scenario in which we extend the minimal model (i.e. one singlet scalar $S_1$ and one VLL) by adding another singlet scalar $S_2$, which along with $S_1$ both play the role of DM. The model is able to explain dark matter and $(g-2)_\mu$ at the EW scale. Multicomponent singlet scalar DM has been an interesting subject to explore \cite{Bhattacharya:2016ysw, Bhattacharya:2017fid, Bhattacharya:2019fgs}, and in particular it has been shown to account for other event excess, such as the cosmic-ray positron-to-electron ratio \cite{Profumo:2019pob}.

The paper is organized in the following way. In section \ref{setup} we introduce the model and its freeze-out mechanism. In section \ref{constraints} we present the most relevant constraints on the model. In section \ref{analysis} we present different scenarios to account for $(g-2)_\mu$ and DM, and finally in section \ref{conclusions} a discussion with the conclusions.

%% file: 2-Model.tex
\section{Set-up}\label{setup}
\subsection{Lagrangian}
We introduce two real singlet scalars $S_1$ and $S_2$, and a fermion $\psi$ being singlet under $SU(2)_L$ with $Y=-1$. Two discrete symmetries are imposed $Z_2\times Z_2'$, where the new fields transform as
\begin{eqnarray}\label{assignments}
 Z_2 &:& S_1, S_2, \psi \rightarrow -S_1, S_2, -\psi , \\
 Z_2' &:& S_1, S_2, \psi  \rightarrow S_1, -S_2, \psi.
\end{eqnarray}
In this way, the Lagrangian is given by
\begin{eqnarray}\label{mainlag}
 \mathcal{L} &\supset & m_\psi \bar{\psi}\psi + (\kappa_1 \bar{\mu}_R S_1 \psi_L + h.c.)   \\ &+&  \sum_{i=1,2}\left(\frac{\mu_i^2}{2}S_i^2 + \frac{\lambda_{Hi}}{2}|H|^2S_i^2 + \frac{\lambda_{i}}{4!}S_i^4\right) + \frac{\lambda_{12}}{4}S_1^2S_2^2 \nonumber
\end{eqnarray}
All the parameters in the model are real. 
For the analysis of this paper the quartic couplings, $\lambda_i$, play no role, so we can set $\lambda_i$ as arbitrary values in our computations.
Additionally, none of the scalar fields get vev, and due to the two discrete symmetries, both $S_1$ and $S_2$ are stable. After electroweak symmetry breaking (EWSB), the masses of both singlet scalars at tree level are given by $m_i^2 = \mu_i^2 + \lambda_{Hi}v_H^2/2$, with $i=1,2$. 

\subsection{Relic abundance}\label{relicsec}
As the model contains two DM components, the total relic abundance is given by the sum of its partial contributions, i.e. $\Omega h^2 = \Omega_1h^2 + \Omega_2h^2$, whose value must be contrasted with the Planck measurement $\Omega_c h^2 = 0.12$ \cite{planck2018}. 
In order to achieve the correct relic abundance and at the same time respecting the new measurements on $(g-2)_\mu$, in our framework it requires to have the following mass hierarchy 
\begin{eqnarray}
 m_2 \lesssim m_1 < m_\psi .
\end{eqnarray}
The first inequality is necessary in such a way that DM conversions of the type $S_1S_1\rightarrow S_2S_2$ be effective in the thermal decoupling history, yielding $\Omega_2 > \Omega_1$, and in this way $\Omega_2$ covering the missing DM in the minimal scenario of two-fields scalar plus VLL \cite{Athron:2021iuf}. The second inequality is necessary in order to have $S_1$ stable, otherwise it could decay into the pair $\psi^\pm\mu^\mp$.

Introducing $x = \mu / T$, with $\mu = m_1m_2/(m_1 + m_2)$ and $T$ the SM temperature, and $Y_i = n_i/s$, the coupled Boltzmann equations for the system of the two singlet scalars in thermal equilibrium in the early universe are given by
	\begin{align}\label{beq_yield}
	\begin{split}
	\dfrac{dY_1}{dx} = &-\lambda_{11XX}\cdot[Y_1^2-Y_{1e}^2]-\lambda_{1122}\cdot
	[Y_1^2-r_{12}^2Y_2^2]\\
	\dfrac{dY_2}{dx} = &-\lambda_{22XX}\cdot[Y_2^2-Y_{2e}^2]+\lambda_{1122}\cdot[Y_1^2-r_{12}^2Y_2^2],
	\end{split}
	\end{align}
with $r_{ij}(x):= Y_{ie}(x)/Y_{je}(x)$ and
	\begin{align}
	\lambda_{iijj}(x):=
	\dfrac{\langle\sigma_{iijj}v\rangle(x)\cdot s(T)}
	{x\cdot H(T)}, \qquad \text{for} ~ i,j=1,2,X.
	\label{def_lambda}
	\end{align}
where $\langle \sigma_{iijj} v \rangle (x)$ is the Maxwell-Boltzmann thermal average annihilation cross section times the relative velocity, $s(T)$ and $H(T)$ the entropy density and Hubble rate, respectively, and $X$ stands for the SM particles, including the Higgs boson (for more details about the Boltzmann equations and its components see \cite{Saez:2021oxl}). 
One important point here is that the Boltzmann equations in eq.~(\ref{beq_yield})
do not include coannihilation processes, e.g. $S_1+\psi^\pm\rightarrow \mu^\pm + V$, with $V = \gamma, Z$, due to the fact that they become effective provided that $m_\psi^2/m_{1}^2 \lesssim 1.2$ \cite{Toma:2013bka}.
We will see in the next section that this mass region is excluded by soft lepton searches, and therefore making coannihilation processes ineffective in the parameter space of our interest.

\begin{figure}[t!]
\centering
\includegraphics[width=0.35\textwidth]{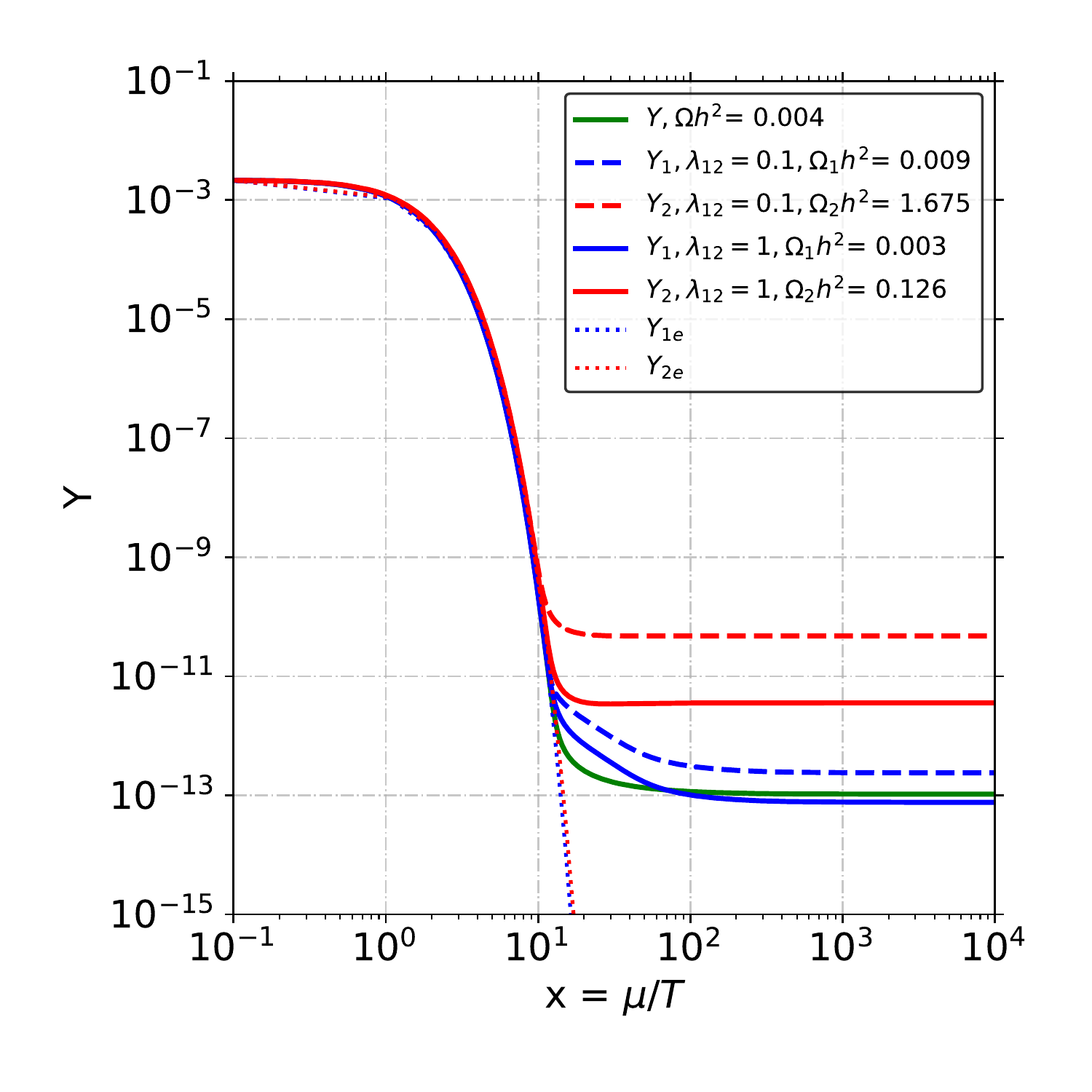}
\caption{Relic yields evolution of $S_1$ (blue curves) and $S_2$ (red curves) as a function of $x$. The parameters are $(m_1, m_2, m_\psi)$ = (130, 125, 150) [GeV], $\lambda_{H1} = \lambda_{H2} = 0$ and $\kappa_1 = 2.8$. The dashed lines consider $\lambda_{12} = 0.1$, whereas for the solid ones $\lambda_{12} = 1$. The green dashed line is the yield of single DM component $S_1$ in a VLL portal, as described in Sec.~\ref{minimal_model}.
}
\label{evo}
\end{figure}

To illustrate the mechanism of DM production (also known as assisted-freeze-out mechanism \cite{Belanger:2011ww, Saez:2021oxl}), in Fig.~\ref{evo} we exemplify the thermal evolution of the yields of the two scalar singlets. We choose the three masses and $\kappa_1$ in the typical ball park to account for $\Delta a_\mu$: $m_\psi = 150$ GeV, $m_1 = 130$ GeV, $m_2 = 125$ GeV and  $\kappa_1 = 2.8$. For simplicity we set $\lambda_{H1} =\lambda_{H2} = 0$. The yields are shown for $\lambda_{12} = 0.1$ (dashed lines) and $\lambda_{12} = 1$ (solid lines). As it can be seen in the plot, after $S_1$ (solid blue lines) decouples from the thermal bath (dotted lines), it continues annihilating into $S_2$ (red lines), also known as DM conversions, reducing its abundance even more. As a consequence, the scenario always predicts $\Omega_1~\ll~\Omega_2$. As a matter of comparison, we show the thermal yield of a single DM scalar in a VLL portal (see Sec.~\ref{minimal_model}) with the green dashed line, making explicit its sub-abundance \cite{Athron:2021iuf}.
One important aspect of this type of mechanism is the relevance of the mass shift $\Delta m \equiv m_1 - m_2$, which must not be bigger than a few GeV \cite{Saez:2021oxl}. As it can be noted from the second line of eq.~(\ref{beq_yield}), 
the relevant term depending on the mass shift is 
\begin{eqnarray}\label{beqdetail}
 \frac{dY_2}{dx} \supset - \lambda_{1122}r_{12}^2Y_2^2 \sim -\lambda_{12}^2e^{-x\frac{\Delta m}{\mu}}Y_2^2.
\end{eqnarray}
This implies that the decreasing rate of $Y_2$ is not only regulated by $\lambda_{12}$, but it also depends strongly on $\Delta m$ and $\mu$. This is, if $\Delta m \rightarrow 0$, there is no exponential suppression, then enhancing the decreasing of $Y_2$, and on the other hand, higher scalar masses (bigger $\mu$) reduces $Y_2$ more efficiently. This last aspect will be part of the analysis in Sec.~\ref{analysis}.

Finally, we solved the system of Boltzmann equations \ref{beq_yield} with our
own code in Python, and cross-checked our results using Micromegas code \cite{Belanger:2013ywg}. 

%% file: 3-Constraints.tex
\section{Constraints}\label{constraints}
On the scenarios in the present work we take into account the following constraints:
\begin{itemize}
 \item \textit{Dark Matter}. First, we set that the total relic predicted by the model $\Omega h^2$ must fulfill the Planck measurement $\Omega_{c}h^2 = 0.12$~\cite{planck2018}, with a tolerance of $\pm$10\% accounting for uncertainties in our computation. Secondly, direct detection (DD) bounds for DM candidates with masses of $\mathcal{O}(100)$ GeV are set by Xenon1T \cite{Aprile_2018}. Two types of portals may be present in the model, and each signal must be weighted by the relative abundance of the corresponding DM component. DM-nucleon interactions via the VLL portal start at two-loops, but this contribution turns out to be well below Xenon1T bounds \cite{Chang:2014tea, Kawamura:2020qxo}. In the case of Higgs portal interactions the spin-independent DM-nucleon scattering amplitude is given by \cite{Cline:2013gha, Hoferichter:2017olk}
 \begin{eqnarray}
  \sigma_\text{SI} = \frac{\lambda_{Hi}^2f_N}{4\pi}\frac{\mu^2 m_n^2}{m_h^4m_i^2}, \quad i=1,2.,
 \end{eqnarray}
where $\mu = m_nm_i/(m_n + m_i)$, $f_N = 0.3$ and $m_n$ the nucleon mass. Finally, the leading indirect detection (ID) signals today are $S_1S_1 \rightarrow \mu\mu (+\gamma)$ or $S_1S_1(S_2S_2)\rightarrow XX$, with $X=Z,W^\pm, b(\bar{b})$ from the VLL and Higgs portal, respectively. We used the most recent Fermi-Lat bounds given in \cite{Bringmann:2012vr}.  
 \item \textit{Collider}. 
 LEP/LHC constrains the mass of the charged VLL to be above $\sim$100 GeV \cite{CMS:2018kag}. Furthermore, the Higgs decay into a pair of muon/antimuon may give rise to important contributions at one-loop level. Considering the measured cross section for the SM Higgs boson production, the upper limit at 95\% C.L. is given by $\text{Br}(h\rightarrow\mu^+\mu^-)~<4.7\cdot~10^{-4}$ \cite{ATLAS:2020fzp, CMS:2020xwi}. Taking into account only $\lambda_{H1}$ coupling (equivalently for $\lambda_{H2}$), the one-loop decay width is given by
 \begin{eqnarray}
  \Gamma(h\rightarrow \mu^+\mu^-) = \frac{\kappa_1^4 \lambda_{H1}^2 v_H^2m_h}{4\pi m_\psi^2}\frac{m_\mu^2}{m_\psi^2}|F(x,y)|^2,
\end{eqnarray}  
with $F(x, y) = \frac{y}{x(y - 1)}\log y + \frac{y - x}{x(x + y +1)}\log(y-x)$, in which $x = m_h^2/m_\psi^2$ and $y \equiv m_1^2/m_\psi^2$. Finally, searches for SUSY long-lived particles and also the CMS searches for the compressed spectra of soft leptons, set limits on the combination $(m_\psi, m_1)$. Here we take the results of \cite{Athron:2021iuf}, where the bounds found there apply directly to our scenario.  
 \item \textit{Flavor Physics}. We take the result in \ref{deltaa} as the reference deviation in the muon anomalous magnetic moment. The leading contribution to $(g-2)_\mu$ in this model comes from the one-loop diagram depicted in Fig.~\ref{gm2}, and it is given by \cite{Hiller:2019mou}
\begin{eqnarray}
 \Delta a_\mu = \frac{\kappa_1^2}{96\pi^2}\frac{m_\mu^2}{m_\psi^2}f\left(\frac{m_{S_1}^2}{m_\psi^2}\right),
\end{eqnarray}
where $f(t) = (2t^3 + 3t^2 - 6t^2\log t - 6t + 1)/(t - 1)^4$, which is positive for any $t$, and $f(0) = 1$. 
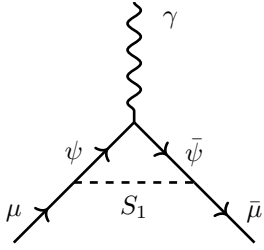
\begin{figure}[t!]
\centering
\begin{tikzpicture}[line width=1.0 pt, scale=0.8]
\begin{scope}[shift={(2,0)}]
	\draw[vector](0,2) -- (0,0);
	\draw[fermionbar](0,0) -- (-1,-1);
	\draw[fermion](0,0) -- (1,-1);
	\draw[scalarnoarrow](-1,-1) -- (1,-1);
	\draw[fermionbar](-1,-1) -- (-2,-2);
	\draw[fermion](1,-1) -- (2,-2);
    \node at (0.6,1.7) {$\gamma$};
	\node at (-1,-0.5) {$\psi$};
	\node at (1,-0.5) {$\bar{\psi}$};
    \node at (-2,-1.5) {$\mu$};
	\node at (2,-1.5) {$\bar{\mu}$};
	\node at (0,-1.4) {$S_1$};
\end{scope}
\end{tikzpicture}
\caption{\footnotesize One-loop contributions to $(g-2)_\mu$ in this work.} \label{gm2}
\end{figure}
\end{itemize}

%% file: 4-Analysis.tex
\section{Analysis}\label{analysis}
In the following we show how the model may account for the correct relic abundance and $(g-2)_\mu$, evading the constraints previously described. Although the parameter space of the model is
\begin{eqnarray}\label{parameters}
 \left(m_\psi,m_1,m_2,\kappa_1, \lambda_{12}, \lambda_{H1}, \lambda_{H2}\right) 
\end{eqnarray}
we start the analysis neglecting the Higgs portal, and afterwards we discuss the effects of it when the portal is not neglected.
\begin{figure}[t!]
\centering
\includegraphics[width=0.35\textwidth]{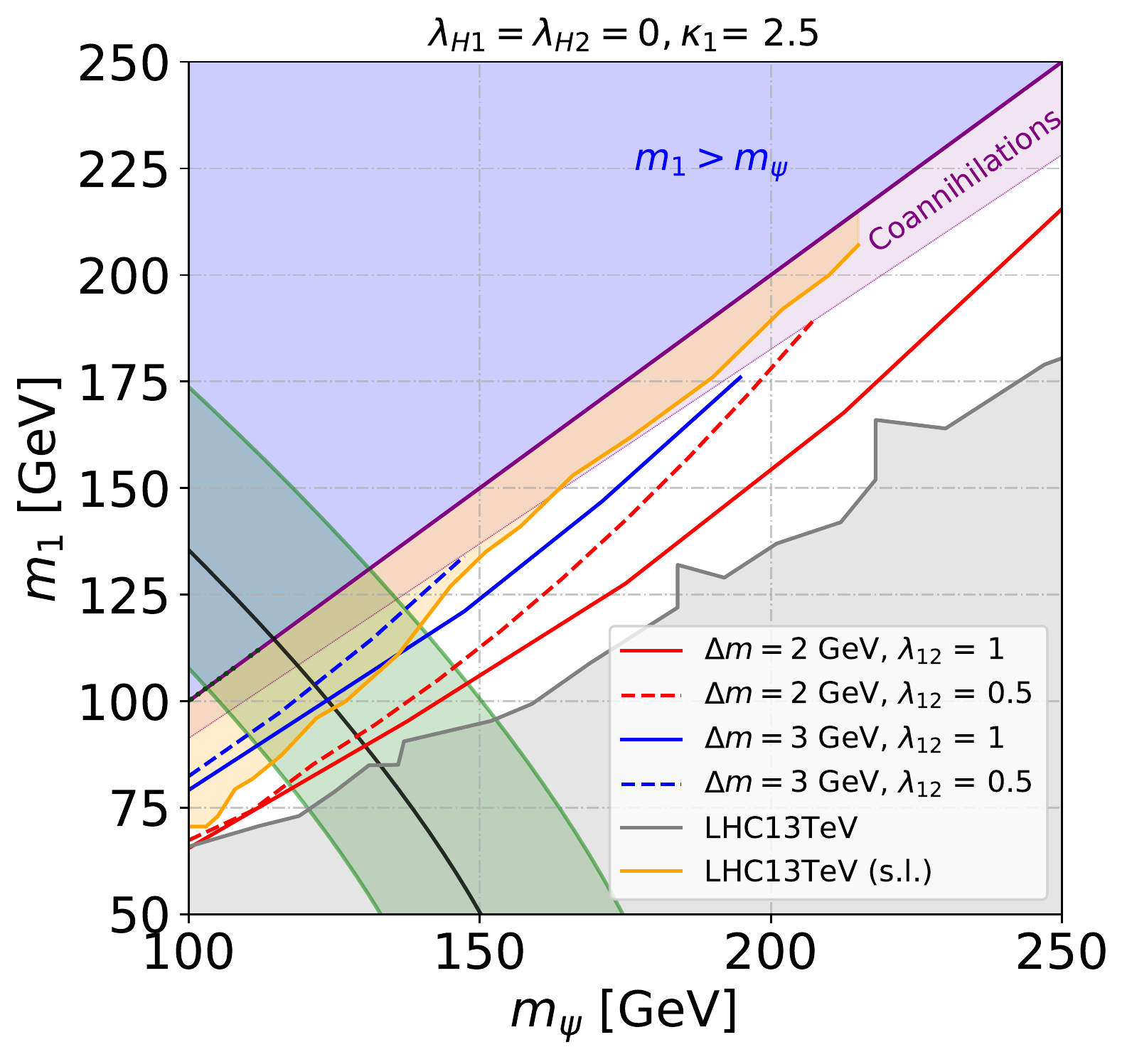}
\includegraphics[width=0.35\textwidth]{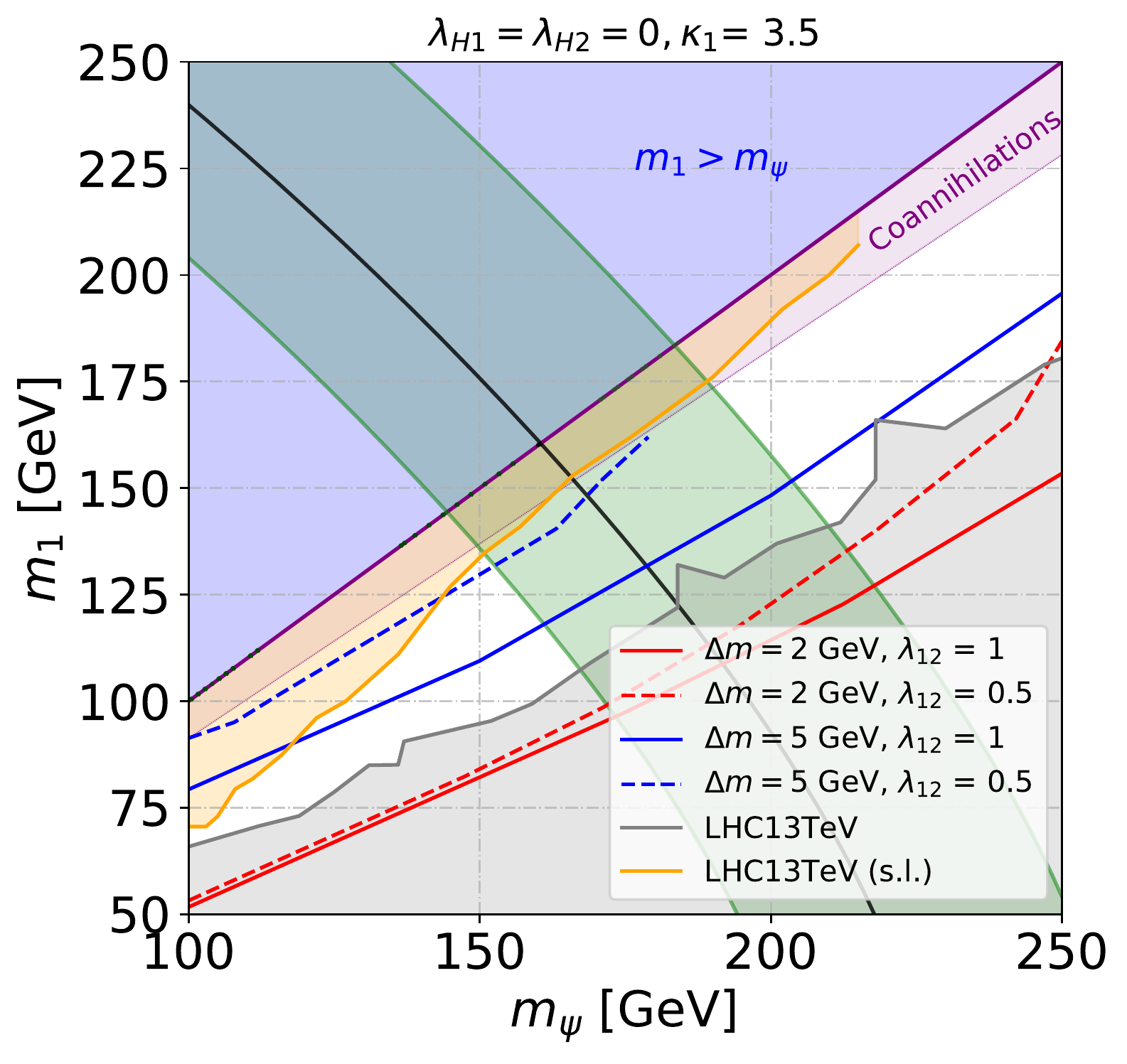}
\caption{Results in the scenario with $\lambda_{H1} = \lambda_{H2} = 0$, scanning over the masses of the new fermion and the heaviest scalar $S_1$, for $\kappa_1 = 2.5$ (plot on the top) and $\kappa_1 = 3.5$ (plot on the bottom). The region that can explain $\Delta a_\mu$ to within 1$\sigma$ is colored green. The blue and red curves indicate those points that produce the correct relic abundance, with the scalar mass difference $\Delta m$ and $\lambda_{12}$ indicated in the legend. The blue region is not allowed by instability of $m_1$, and the yellow and grey regions are LHC 13 TeV and soft leptons (s.l.) searches obtained in \cite{Athron:2021iuf} (for details see the experimental references therein).} 
\label{random1}
\end{figure}
\subsection{Higgs portals closed}\label{simplestscenario}
We start assuming that the Higgs portals are sub-dominant, i.e. $(\lambda_{H1},\lambda_{H2})\ll 1$. In this case, the model becomes very simple since only the first four parameters of \ref{parameters} are relevant, and the constraints presented in the last section reduce to a few ones: relic abundance, $(g-2)_\mu$ and some collider constraints. Then, for simplicity we assume that $\lambda_{H1} =\lambda_{H2}=0$. 

In Fig.~\ref{random1} we show some predictions of the model projected on the plane $(m_\psi, m_1)$ in which all the constraints are fulfilled and $(g-2)_\mu$ is accounted, for $\kappa_1 = 2.5$ (plot on the top) and $\kappa_1 = 3.5$ (plot on the bottom). The white region is still allowed by the existing constraints, whereas the yellow and grey regions are already ruled out by the collider bounds. The blue region is not allowed in this model because it would make $S_1$ unstable ($S_1 \rightarrow \psi^\pm + \mu^\mp$), and the purple one is where coannihilations become effective. The green band accounts for $(g-2)_\mu$ at 1$\sigma$, with the black line near the middle corresponds to the central value. 

As it can be noted from the plot in the upper panel of Fig.~\ref{random1}, $m_\psi$ and $m_1$ spanning several dozens of GeV with small $\Delta m$ and $\lambda_{12}\approx 1$ enter the region allowed by the experimental constraints fulfilling $(g-2)_\mu$. In particular, red curves with $\Delta m = 2$ GeV enter the green zone, with tiny deviations as $\lambda_{12}$ changes, and as $\Delta m$ increases the model predictions deviate from the allowed region, as it can be noted for $\Delta m = 3$ GeV (blue curves). The strong sensitivity of the model predictions to the growing of $\Delta m$ can be understood due to the fact that in order to keep the correct relic abundance at higher $\Delta m$, it is necessary to increment the reduced mass $\mu$ to compensate the exponential suppression given in eq.~(\ref{beqdetail}). The sensitivity of the model predictions to $\lambda_{12}$ peaks at high masses where the exponential suppression in eq.~(\ref{beqdetail}) is less strong, therefore making the abundance $Y_2$ being sensitive mainly to $\lambda_{12}^2$.
Equivalently, the plot in the lower panel of Fig.~\ref{random1} considering $\kappa_1 = 3.5$ shows that the viable mass shift picks up a little bigger value, as it is illustrated for $\Delta m = 5$ GeV (blue curves), with an increasing range for the masses $m_\psi$ and $m_1$. Note that as a big portion of the coannihilation region is excluded by LHC bounds (yellow region) and it practically does not match the $(g-2)_\mu$ region (green one) we did not show the predictions of the model in that region.


In this simple scenario, as the Higgs portal is closed, the leading contributions to ID are $S_1S_1\rightarrow\mu^+\mu^-, \mu^+\mu^-\gamma,\gamma\gamma$. However, in our scenarios $S_1$ is always a sub-dominant DM component, typically $\Omega_1 h^2 \sim 10^{-3}$, then the resulting re-scaled average annihilation cross section at zero velocity is suppressed by a factor of order $\sim 10^{-6}$. We have checked that the resulting signals are some order of magnitude below the upper bounds from Fermi-LAT.

\subsection{Higgs portals opened}
\begin{figure}[t!]
\centering
\includegraphics[width=0.35\textwidth]{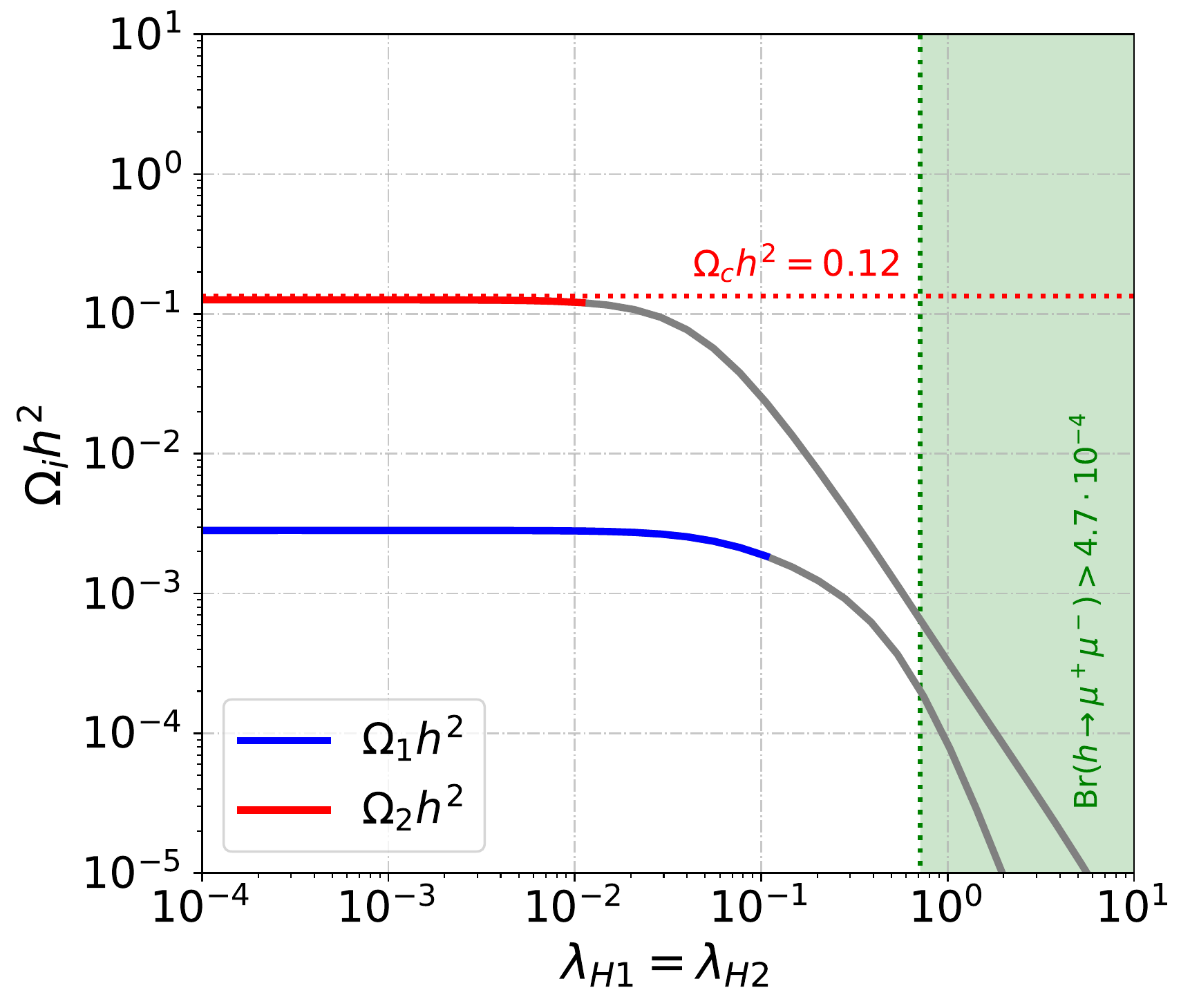}
\includegraphics[width=0.35\textwidth]{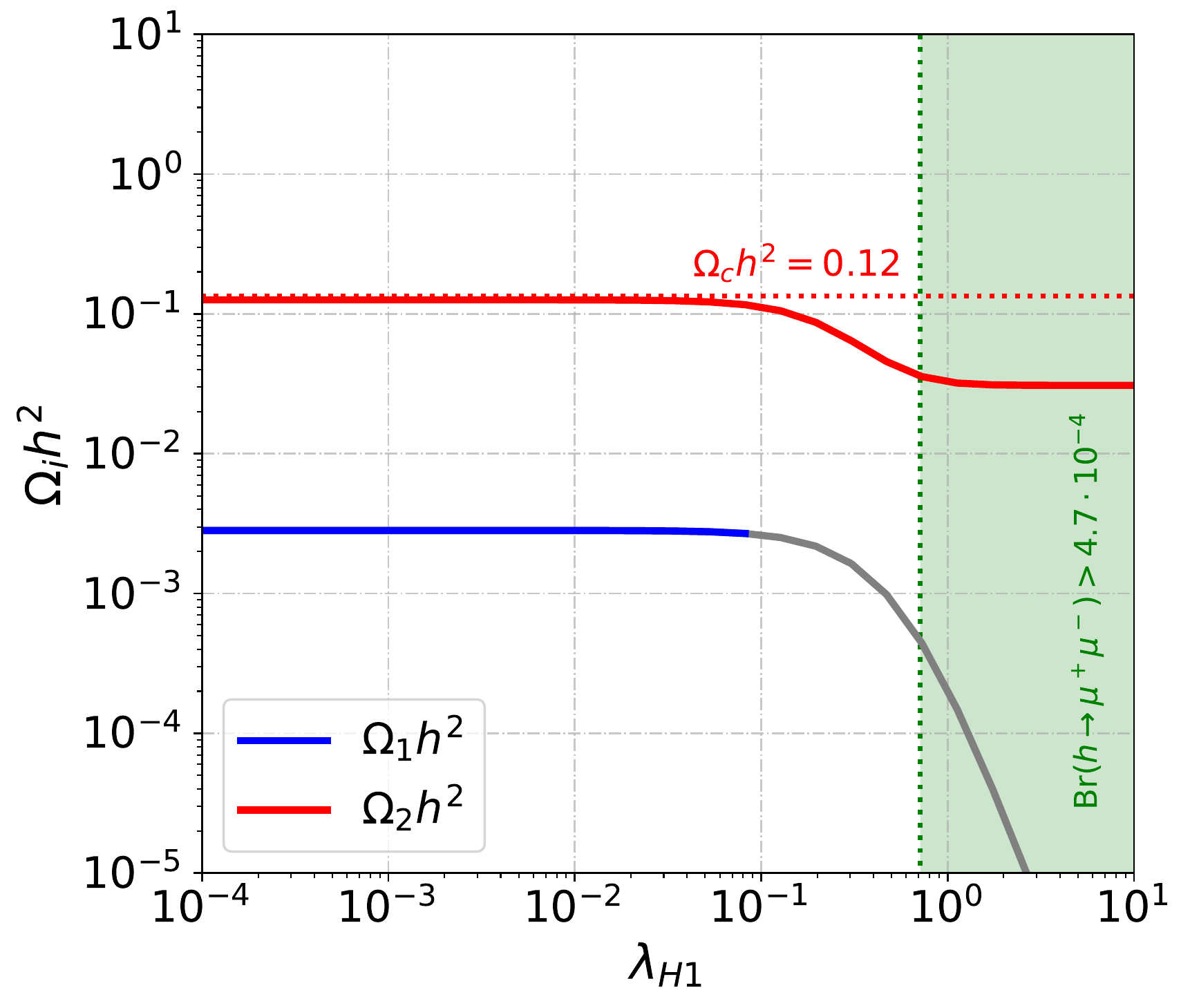}
\caption{Relic abundances as a function of Higgs portal parameters, with $(m_\psi, m_1, m_2) =(150, 130, 125)$ GeV, $\kappa_1 = 2.8$ and $\lambda_{12} = 1$. In the plot above we consider $\lambda_{H1} = \lambda_{H2}$, whereas in below $\lambda_{H2} = 0$. The blue and red curves are allowed by DD bounds, whereas the grey one is the already ruled out by Xenon1T. The green region is forbidden by Br$(h\rightarrow \mu^+\mu^-)$ measurements, and the dotted horizontal line is the measured relic abundance $\Omega_c h^2$.} 
\label{omelh1}
\end{figure}

Now we consider the case that either one or two Higgs portals are not negligible. In this way, here we sketch what deviations occur once the Higgs portals are not negligible with respect to the previous scenario. Let us start assuming that $\lambda_{H1} = \lambda_{H2} \neq 0$. It is not possible to arbitrarily increase the values of the Higgs portal, because they not only will deviate the relic abundance found in the previous scenario, but they increase the DD signal. Without loss of generality, let us consider a point of the parameter space that fulfills $(g-2)_\mu$ and LEP/LHC constraints: $(m_\psi, m_1, m_2) =(150, 130, 125)$ GeV, $\kappa_1 = 2.8$ and $\lambda_{12} = 1$. In Fig.~\ref{omelh1}(upper panel) we show the deviation in the relic abundances as the Higgs portal couplings change according to $\lambda_{H1} = \lambda_{H2}$. 
For $\lambda_{H2} \gtrsim 0.01$, modification on $\Omega_2$ starts appearing with respect to the case $\lambda_{H1} = \lambda_{H2} = 0$
and almost simultaneously the model enters the region in conflict with the DD bounds (grey curve).
Therefore, provided that the Higgs portal couplings take values lower than 
$\mathcal{O}(0.01)$, no significant deviations with respect to what we have found in Sec.~\ref{simplestscenario} will appear in this case. 


Since the scalar $S_2$ gives most of the DM relic abundance (i.e. in this scenario we always have $\Omega_1 \ll \Omega_2$) and therefore most sensible to the DD bounds via the Higgs portal, another interesting option is to set $\lambda_{H1} \neq 0$ keeping $\lambda_{H2}$ very small. In this case, higher values for $\lambda_{H1}$ can be reached before the DD bounds start to be relevant. In particular, for the same parameter space point described in the previous paragraph, in Fig.~\ref{omelh1}(lower plot) $\lambda_{H1} \lesssim 0.1$ still fulfills the relic abundance while evading DD bounds. Therefore, allowing the Higgs portals coupling take non-vanishing values, DD bounds kill all the significant deviations to the predictions obtained in Sec.~\ref{simplestscenario} (Higgs portal closed).

As a matter of complement, note that the scenario $\lambda_{H1}\neq 0$ and $\lambda_{H2} = 0$ is not unstable under radiative corrections due to the following. At one-loop, the effective $S_2S_2h$ coupling in the limit in which $\lambda_{H2} \ll 1$ is given by \cite{Saez:2021oxl}
\begin{eqnarray}\label{effvertex1}
 \lambda_{H2}^\text{eff} = \lambda_{H2} + \frac{\lambda_{H1}\lambda_{12}}{16\pi^2}. 
\end{eqnarray}
As we have seen, the maximum value of $\lambda_{H1}$ allowed by the DD bounds is $\mathcal{O}(0.1)$, then implying that at one-loop $\lambda_{H2}^\text{eff} \approx 6.3 \times 10^{-4}$, for $\lambda_{12} \sim \mathcal{O}(1)$. As we have seen in this section, such small values for $\lambda_{H2}^\text{eff}$ produce no deviation on the relic abundance and evade easily the DD bounds.

%% file: 5-Conclusions.tex
\section{Discussion and Conclusions}\label{conclusions}

In this work we have shown that a simple unexplored extension to the SM containing two real singlet scalars and one vector-like lepton not only accounts for the measured DM relic abundance evading stringent experimental constraints, but it can also fulfill the recent $4.2\sigma$ deviation in $(g-2)_\mu$, to be tested in the forthcoming years with more data. Provided that the mass of the new three states be around the Fermi scale and the mass shift between the singlet scalars is not more than a few GeV, the $(g-2)_\mu$ anomaly is perfectly achieved, with the minimal requirement that the new couplings be of order one. We have seen that this set-up implies that the inert scalar $S_2$ captures most of the relic abundance, whereas the sub-leading component $S_1$ plays the role of enhancing the anomalous magnetic moment of the muon at one-loop. Furthermore, we have seen that the Higgs portal couplings play a secondary role in this framework, although they can still take relatively sizable values in agreement with all the experimental constraints. Finally, an additional comment about the possibility that also $S_2$ couples to the muon via the VLL such that both scalars participate in the $(g-2)_\mu$ (this can be done assuming $\psi\rightarrow -\psi$ under both $Z_2$ and $Z_2'$). In that case, one can have $\kappa_1$ and $\kappa_2$ with slightly lower values than the obtained ones in this work. That scenario is dangerous because the scalar with the highest DM relic abundance (i.e. at least half of $\Omega_c h^2$) will present strong indirect signals (e.g. bremsstrahlung), then entering in conflict with the present ID bounds. 
\section{Acknowledgment}
B.D.S has been founded by ANID (ex CONICYT) Grant No. 74200120.